\documentclass[prb,showpacs,superscriptaddress,prepreint,twocolumn]{revtex4}

\usepackage{graphicx}
\usepackage{dcolumn}
\usepackage{bm}

\begin{document}

\title{The 0 and the $\pi$ phase Josephson coupling through an insulating barrier with magnetic impurities}

\author{O. V\'{a}vra}
\email{ondrej.vavra@savba.sk} \altaffiliation[Present address:
]{Institut f\"{u}r experimentelle und angewandte Physik,
Universit\"{a}t Regensburg, D-93025 Regensburg, Germany}
\affiliation{INPAC-Institute for Nanoscale Physics and Chemistry,
Nanoscale Superconductivity and Magnetism Group, K. U. Leuven,
Celestijnenlaan 200 D, B-3001 Leuven, Belgium}
\affiliation{Institute of Electrical Engineering, Slovak Academy of
Sciences, D\'{u}bravsk\'{a} cesta 9, SK-841 04 Bratislava, Slovak
Republic}

\author{S. Ga\v{z}i}
\affiliation{Institute of Electrical Engineering, Slovak Academy
of Sciences, D\'{u}bravsk\'{a} cesta 9, SK-841 04 Bratislava,
Slovak Republic}

\author{D. S. Golubovi\'{c}}
\affiliation{INPAC-Institute for Nanoscale Physics and Chemistry, Nanoscale Superconductivity and
Magnetism Group, K. U. Leuven, Celestijnenlaan 200 D, B-3001 Leuven, Belgium}

\author{I. V\'{a}vra}
\affiliation{Institute of Electrical Engineering, Slovak Academy
of Sciences, D\'{u}bravsk\'{a} cesta 9, SK-841 04 Bratislava,
Slovak Republic}

\author{J. D\'{e}rer}
\affiliation{Institute of Electrical Engineering, Slovak Academy of Sciences, D\'{u}bravsk\'{a}
cesta 9, SK-841 04 Bratislava, Slovak Republic}

\author{J. Verbeeck}
\affiliation{EMAT, University of Antwerp, Groenenborgerlaan 171, B-2020, Antwerp, Belgium}

\author{G. Van Tendeloo}
\affiliation{EMAT, University of Antwerp, Groenenborgerlaan 171, B-2020, Antwerp, Belgium}

\author{V. V. Moshchalkov}
\affiliation{INPAC-Institute for Nanoscale Physics and Chemistry, Nanoscale Superconductivity and
Magnetism Group, K. U. Leuven, Celestijnenlaan 200 D, B-3001 Leuven, Belgium}

\date{\today}

\begin{abstract}
We have studied temperature and field dependencies of the critical current $I_{C}$ in the
\hbox{Nb-Fe$_{0.1}$Si$_{0.9}$-Nb} Josephson junction with tunneling barrier formed by paramagnetic
insulator. We demonstrate that in these junctions the co-existence of both the 0 and the $\pi$
states within one tunnel junction takes place which leads to the appearance of a sharp cusp in the
temperature dependence $I_{C}(T)$ similar to the $I_{C}(T)$ cusp found for the $0-\pi$ transition
in metallic $\pi$ junctions. This cusp is not related to the $0-\pi$ temperature induced transition
itself, but is caused by the different temperature dependencies of the opposing 0 and $\pi$
supercurrents through the barrier.
\end{abstract}


\maketitle As first predicted by Josephson \cite{Josephson1962}, the supercurrent $I_{S}$ through
the tunnel barrier is driven by the phase difference $\varphi$ across the junction applied to the
superconducting wave function. In conventional Josephson junctions (JJ) this current is described
by the relation \hbox{$I_{S}=I_{C}sin \varphi$}, where $I_{C}$ is the critical current. Recently, a
considerable attention has been devoted to the investigation of the $\pi$ JJs
\cite{Kontos2002,Ryazanov2001,Ryazanov2004}. In this case the relation between the supercurrent and
the phase difference is \hbox{$I_{S}=I_{C}sin (\varphi+\pi)=-I_{C}sin \varphi$}
\cite{Bulaevskii1977}. One of the possible realizations of the $\pi$ junctions is the
superconductor-ferromagnetic metal-superconductor (\hbox{S-FM-S}) tunnel junction, wherein the
spatial oscillations of the superconducting order parameter occur in the ferromagnetic metal as a
consequence of the exchange splitting of the conduction band \cite{Buzdin1982}. The transition
between the 0 and the $\pi$ states was experimentally observed as the vanishing of the Josephson
current. The \hbox{$0-\pi$} transition can be induced by the varying barrier thickness
\cite{Kontos2002,Ryazanov2004} or the temperature \cite{Ryazanov2001,Ryazanov2004}. As the absolute
value of the current is measured, e.g. for proper values of the ferromagnetic barrier thickness, a
sharp cusp in the temperature dependence of the critical current $I_{C}(T)$ is observed as a
consequence of the $0-\pi$ transition.

It is also predicted that JJs with magnetic impurities within an \textit{insulating barrier} can
produce the $\pi$ state \cite{Bulaevskii1977}. Later on, the possibilities to observe the $\pi$
junctions in JJs with ferromagnetic insulating or semiconducting barrier (\hbox{S-FI-S}) were
analyzed theoretically in \cite{Fogelstrom2000,Barash2002}. In such types of JJs the proximity
effect in the barrier is much weaker, as compared with the ferromagnetic metal, and can be
disregarded. In this case the formation of the $\pi$ junction is caused by the quasiparticle
scattering on a magnetically active interfaces \cite{Fogelstrom2000}. It can result in the
splitting of the Andreev interface bound-state energies into two spin channels \cite{Barash2002}.
Theoretically, if these channels compensate each other the \hbox{$0-\pi$} transition is observed.
Up to now the $\pi$ state in the JJs with insulating magnetic barrier has not been found
experimentally.

\begin{figure}[b]
\centering
\includegraphics*[width=8.6 cm]{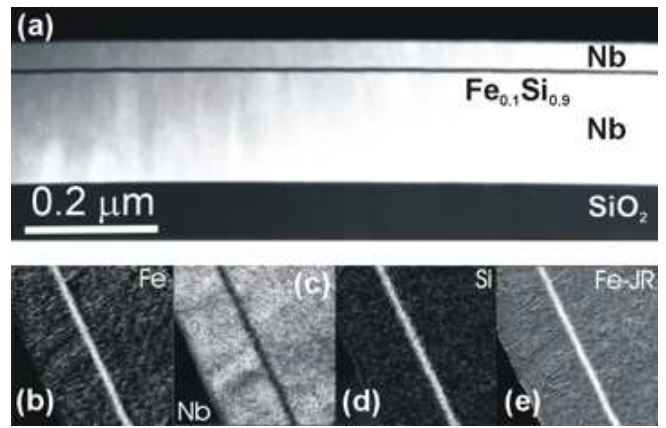}
\caption{Cross-sectional micrographs of the \hbox{Nb-Fe$_{0.1}$Si$_{0.9}$-Nb} tunnel junction
obtained in various modes of transmission electron microscope: (a) z-contrast annular dark-field
micrograph, (b),(c) and (d) energy filtered TEM elemental maps of Fe, Nb and Si, (e) jump-ratio Fe
elemental map. The thickness of the \hbox{Fe$_{0.1}$Si$_{0.9}$} barrier is \hbox{6 nm}.\label{TEM}}
\end{figure}

In this letter we present the experimental evidence of the existence of the $\pi$ state in the
Josephson junction with magnetic impurities in the insulating barrier. We also demonstrate that the
co-existence of both the 0 and the $\pi$ states within one tunnel junction leads to the appearance
of a similar cusp in the temperature dependence of $I_{C}$ as the one for the $0-\pi$ transition in
metallic $\pi$ junctions. The origin of this cusp, however, is not related to the $0-\pi$
transition itself, but rather to a simultaneous 0 and $\pi$ Josephson tunneling through an
insulating barrier with magnetic impurities. In this case, quite generally, the tunneling through
an insulating barrier itself gives rise to a positive contribution into the critical current
$I_{C0}$ (0-part), whereas the tunneling via scattering on magnetic impurities generates a negative
$I_{C\pi}$ ($\pi$-part). These two currents have opposite signs and different temperature
dependencies of the critical currents, which results in their complete mutual cancelation
$|I_{C0}-I_{C\pi}|=0$ at a certain temperature where a sharp depression of the critical current has
been observed.

\begin{figure}[t]
\centering
\includegraphics*[width=8cm]{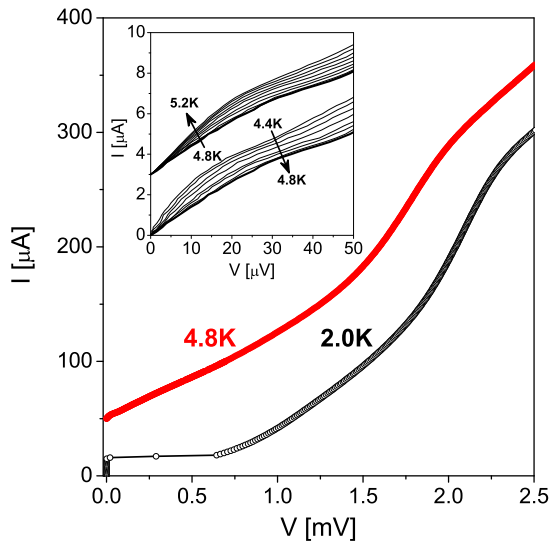}
\caption{Current-voltage characteristics of the \hbox{Nb-Fe$_{0.1}$Si$_{0.9}$-Nb} tunnel junction
at \hbox{4.8 K} and \hbox{2.0 K}. The curves are shifted by \hbox{$50\mu A$} for clarity. Inset:
The temperature dependent change of the slope of the IV curves around \hbox{4.8 K}. The curves for
the range from \hbox{4.8 K} to \hbox{5.2 K} are shifted by $3\mu V$ for clarity.}\label{IV}
\end{figure}

\begin{figure}[t]
\centering
\includegraphics*[width=8cm]{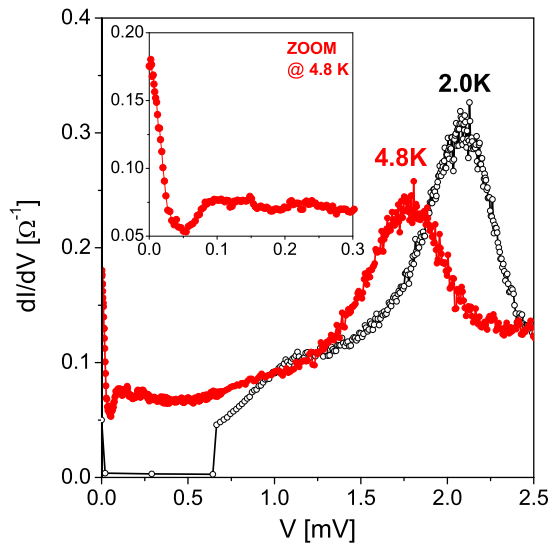}
\caption{Differential conductance vs. bias voltage characteristics of the
\hbox{Nb-Fe$_{0.1}$Si$_{0.9}$-Nb} junctions at \hbox{4.8 K} and \hbox{2.0 K}. Inset: Zoom, showing
the zero-bias conductance peak at \hbox{4.8 K.}\label{dIdV}}
\end{figure}

We have studied a \hbox{Nb-Fe$_{0.1}$Si$_{0.9}$-Nb} tunnel junction, with the amorphous
\hbox{Fe$_{0.1}$Si$_{0.9}$} alloy as the barrier. Amorphous magnetic materials here have certain
advantages compared to polycrystalline materials, because of a lack of crystalline defects and a
better composition homogeneity at a microscopic level. Moreover, \hbox{Fe$_{0.1}$Si$_{0.9}$} alloy
is additionally favorable, since it is an insulator at low temperatures, with the resistance a few
orders of magnitude higher than that of the metallic alloys. The \hbox{Fe$_{0.1}$Si$_{0.9}$} alloy
is a paramagnetic material \cite{OVavra2002}, but the amorphous structure does not rule out
completely the possible existence of a local ferromagnetic exchange field at low temperatures. The
molecular dynamics ab-initio simulation reveals that nearest neighbor (NN) positions of the Fe
atoms are also quite probable. From that point of view, the formation of microscopic regions with
ferromagnetic exchange coupling seems to be possible. In this case the barrier can be thought of as
a "nanocomposite", containing regions with and without magnetic exchange coupling.

The \hbox{Nb-Fe$_{0.1}$Si$_{0.9}$-Nb} junctions were prepared by a sputtering technique under the
conditions similar to those used for the fabrication of the \hbox{Nb-Si-Nb} junctions
\cite{Lobotka1997}. The area of the junction is $20 \times 20\mu m^{2}$. The structure and
composition homogeneity of the 6 nm thick barrier were investigated by transmission electron
microscopy (TEM) in a cross-sectional specimen (fig.~\ref{TEM}). These studies have clearly
confirmed that the barrier is very well defined. Indications of neither strong interdiffusion nor
local Nb shorts were found. The \hbox{Fe$_{0.1}$Si$_{0.9}$} barrier is very homogeneous in
thickness as well as in composition.

The current-voltage characteristics (IV) of the \hbox{Nb-Fe$_{0.1}$Si$_{0.9}$-Nb} junction were
measured (Fig.~\ref{IV}), and subsequently the differential conductance ($dI/dV$) versus the bias
voltage was determined numerically (Fig.~\ref{dIdV}). At both temperatures 4.8 K and 2 K peaks are
observed at voltages \hbox{$V = 1.77 mV$} and \hbox{$V = 2.09 mV$}, respectively. They correspond
to the sum of the superconducting gaps related to the individual Nb electrodes of the tunnel
junction. A reduced value of the sum of the superconducting gaps is most probably due to a
non-ideal upper Nb electrode (see below).

At the temperature of 2K, the critical current is \hbox{$I_{C}=17 \mu A$}. As the $I_{C}$ value is
finite, the derivative of the current-voltage curve at the zero bias is infinite. Such $IV$ and
$dI/dV$ curves are typical for the temperatures and magnetic fields where the junction has a finite
$I_{C}$ value.

\begin{figure}[t]
\centering
\includegraphics*[width=8cm]{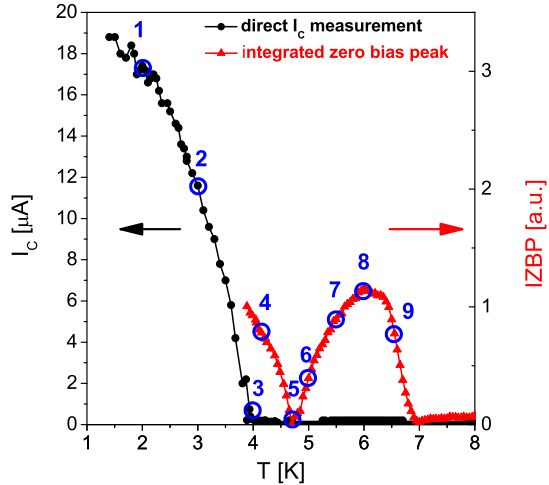}
\caption{The temperature dependencies of the critical current $I_{C}(T)$ and integrated zero-bias
peak $IZBP(T)$ of the \hbox{Nb-Fe$_{0.1}$Si$_{0.9}$-Nb} tunnel junction in the zero magnetic field.
The $IZBP(T)$ can be considered as a zoom of the $I_{C}(T)$ dependence in the temperature range
from \hbox{4 K} up to \hbox{7 K}. The circles marks denotes the points where the $I_{C}(\Phi)$ was
measured [see Fig.~\ref{IcB}].\label{IcT}}
\end{figure}

\begin{figure}[t]
\centering
\includegraphics*[width=8cm]{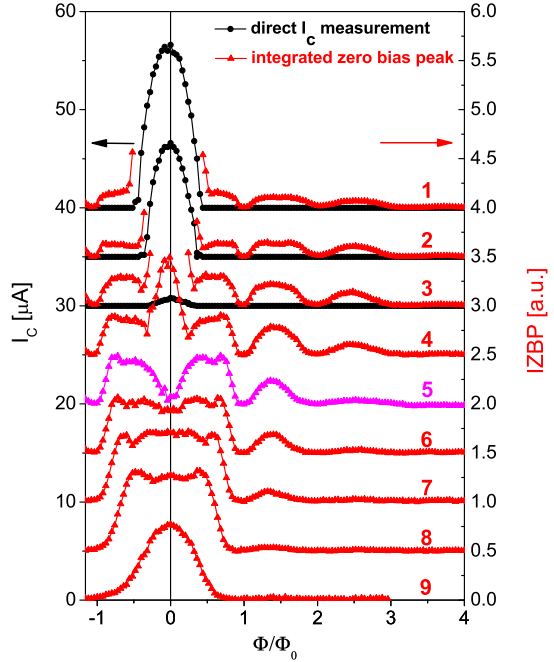}
\caption{The applied magnetic flux dependencies of the critical current $I_{C}(\Phi)$ and
integrated zero-bias peak $IZBP(\Phi)$ of the \hbox{Nb-Fe$_{0.1}$Si$_{0.9}$-Nb} tunnel junction for
temperatures marked in the Fig.~\ref{IcT}.\label{IcB}}
\end{figure}

The $dI/dV$ curve measured at \hbox{4.8 K} corresponds to the applied magnetic flux
$\Phi/\Phi_{0}=0.7$ where the maximum of the zero-bias peak was observed [see Fig.~\ref{IcB}]. Such
type of the $dI/dV$ curves is typical for the temperatures and magnetic fields where the measurable
critical current is absent.

For the reference junction \hbox{Nb-Si-Nb} \cite{Lobotka1997} the upper Nb electrode contains a
thin (\hbox{$\sim$ 2 nm}) sublayer of amorphous Nb adjacent to the silicon barrier. This junction
behaves like the SINS system, where N represents the amorphous part of the Nb electrode. Due to the
non-equal atomic condensation of Nb on Fe$_{0.1}$Si$_{0.9}$, similar amorphous Nb sublayer was also
found in the \hbox{Nb-Fe$_{0.1}$Si$_{0.9}$-Nb} junctions. The absence of the measurable critical
current at zero bias is caused by the decay of the superconducting order parameter in this
amorphous part of the Nb electrode as well as by the Andreev scattering at the interface of the
polycrystalline and amorphous Nb. Similar zero-bias peak in dI/dV was described by Klapwijk
\cite{Klapwijk1994} and in his case of the Nb-Si-Nb junction was considered as a precursor of the
fully developed supercurrent observed for thinner barriers. In our case the zero-bias peak in dI/dV
can be interpreted as a precursor of the supercurrent for thinner amorphous part of the upper Nb
electrode. To obtain the information about the precursor of the $I_{C}$ we introduced the
integrated zero-bias peak ($IZBP$) amplitude

\begin{equation} \label{eq:Integral}
    IZBP = \int_0^{V_C}(dI/dV(V)-dI/dV_{offset}) dV,
\end{equation}
where $V_C$ is a voltage criterion (we used \hbox{$V_C = 5 \mu V$}), and $dI/dV_{offset}$ is the
reference conductance value. From the $IZBP(T)$ and $IZBP(\Phi)$ data (see below) we confirm that
the zero-bias peak is the precursor of $I_{C}$.

Direct measurements of the $I_{C}$ at the zero magnetic field reveal some finite values up to the
temperature around \hbox{4 K} [see Fig.~\ref{IcT}]. Above this temperature, instead of $I_{C}$ the
zero bias-peak is observed in the $dI/dV$ curves. To find out what happens above 4 K, we have
determined the $IZBP$ for each used temperature $T>4 K$. As it can be seen from Fig.~\ref{IcT}, the
proposed method reveals a sharp $IZBP$ cusp at the temperature of \hbox{4.8 K} [for correlation see
also the temperature dependent change of the slope of the IV curves around the zero bias in the
inset in Fig.~\ref{IV}], the maximum at approximately \hbox{6 K}, and then decrease down to zero at
$T\approx7K$ which is the critical temperature of our \hbox{Nb-Fe$_{0.1}$Si$_{0.9}$-Nb} JJ.

Fig.~\ref{IcB} shows the critical current vs. applied magnetic flux $I_{C}(\Phi)$ dependencies for
the temperatures marked in Fig.~\ref{IcT} by circles. Curves 1, 2 and 3 show the peak around the
zero field with finite values of $I_{C}$. Elsewhere, the $I_{C}(\Phi)$ dependence is suppressed and
was, therefore, obtained using $IZBP$ versus the applied magnetic flux. Similarly to $IZBP$ versus
the temperature, the $IZBP$ was found for each value of the magnetic flux. In what follows the
$IZBP$ together with the $I_{C}$ temperature and magnetic field dependencies will be referred to as
a single $I_{C}(T)$ or $I_{C}(\Phi)$ dependence.

The unusual behavior of the junction in magnetic field is clearly seen from Fig.~\ref{IcB}. As the
temperature decreases, the shape of the $I_{C}(\Phi)$ changes. Especially, in the interval of
magnetic flux \hbox{$\Phi\in\langle-\Phi_{0}; \Phi_{0}\rangle$} it is visible that the middle peak
gradually vanishes as the temperature increases (curves 1-4). At the temperature of 4.8 K (curve 5)
the minimum of the critical current at zero applied flux $I_{C}(0)$ is observed. Such behavior with
the minimum of critical current at $\Phi = 0$ is typical for the $0-\pi$ JJs \cite{Bulaevskii1978}.
Then for the temperatures \hbox{$4.8 K < T < 6 K$} the shape of the $I_{C}(\Phi)$ curves is
changing again and the $I_{C}(0)$ recovers its zero field maximum (curves 6-9). It is worth noting
that in a reference Nb-Si-Nb JJ the $I_{C}(\Phi)$ shows a well defined conventional Fraunhofer like
patterns \cite{Vavra1994}

\begin{figure}[tb]
\centering
\includegraphics*[width=8cm]{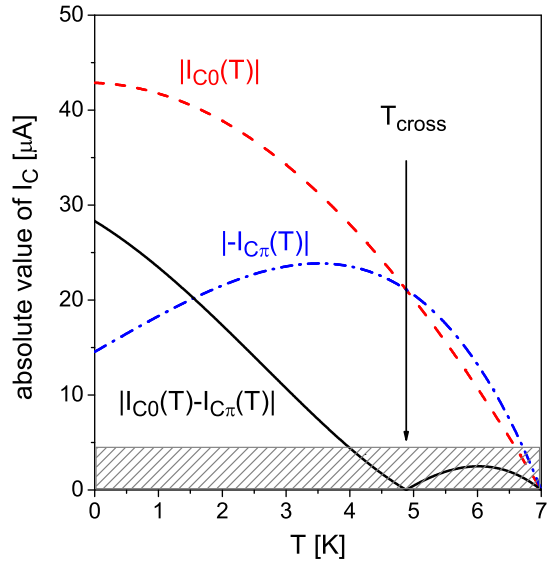}
\caption{Simplified illustrative model  of the
\hbox{Nb-Fe$_{0.1}$Si$_{0.9}$-Nb} junction: The theoretical partial
temperature dependencies of the critical current for the 0- and
$\pi$-parts and their sum, showing the cusp in the \hbox{$I_{C}(T)$}
caused by crossing of the \hbox{$I_{C0}(T)$} and
\hbox{$I_{C\pi}(T)$} curves. Shaded area: The levels of the $I_{C}$
measured by IZBP due to the absence of the measurable critical
current.\label{IcTtheory}}
\end{figure}

As predicted Bulaevskii \cite{Bulaevskii1978} a perfectly flat and homogeneous insulating barrier
with magnetic impurities can induce the formation of an admixture of the $\pi$ and 0 junctions
("vortex states") for certain range of the barrier parameters and temperatures. Since the $0-\pi$
phase boundary corresponds to the nucleation of a semifluxon, this "vortex phase" is, in fact a
collection of semifluxons formed at the $0-\pi$ barrier boundaries. Another possible reason for the
co-existence of the $\pi$ and 0 phases could be the barrier thickness modulation or/and formation
of the Fe clusters. However, taking into account a very homogeneous and flat boundaries
(Fig.~\ref{TEM}), a pure paramagnetic behaviour and a lack of electron diffraction rings typical
for nanocristallites in an amorphous matrix, the latter scenario seems to be less probable.

In our case the co-existence of the 0 and $\pi$ junctions can be simulated as a JJ with a
nonuniform spatial distribution of the critical current density and with additional polarity
alternations. The assumption about the simultaneous presence of the 0 and the $\pi$ tunneling is
confirmed by the unusual shape of the $I_{C}(\Phi)$ curves (Fig.~\ref{IcB}). When both 0 and $\pi$
phases of the Josephson supercurrent co-exist in one \hbox{Nb-Fe$_{0.1}$Si$_{0.9}$-Nb} junction,
two $I_{C}(T)$ dependencies must be taken into account: $I_{C0}(T)$ and $I_{C\pi}(T)$
(Fig.~\ref{IcTtheory}). Due to the lack of the adequate theories for such kind of the junctions the
$I_{C0}(T)$ and the $I_{C\pi}(T)$ dependencies were calculated by using the theory of the $\pi$
junctions with metallic barrier \cite{Buzdin2005,Ryazanov2004}. In this illustrative simulation the
$I_{C0}(T)$ and the $I_{C\pi}(T)$ are taken for the same barrier thickness but with different
values of the ferromagnetic exchange energy, the decay length $\xi_{F1}$, and oscillation period of
the order parameter $2\pi\xi_{F2}$ (details will be provided elsewhere). The sum of these two
currents of the opposite polarities (positive for $I_{C0}$ and negative for $I_{C\pi}$) gives the
$I_C(T)$ dependence which is similar to the one we have found (compare Fig.~\ref{IcT} and
Fig.~\ref{IcTtheory}). The minimum of the $I_C(T)$ dependence $T_{cross}$ corresponds to the
crossing point of the $|-I_{C\pi}(T)|$ and $|I_{C0}(T)|$ dependencies [see Fig.~\ref{IcTtheory}].

In conclusion, we have observed the co-existence of the 0 and the $\pi$ state in the
\hbox{Nb-Fe$_{0.1}$Si$_{0.9}$-Nb} Josephson junctions with a paramagnetic insulating barrier formed
by an amorphous Fe$_{0.1}$Si$_{0.9}$. Different temperature dependencies of the $I_{C0}$ and
$I_{C\pi}$ currents and their opposite signs lead to the appearance of very sharp cusp in the
$I_{C}(T)$ curve at about 4.8 K where these two currents cancel each other completely. The
simultaneous presence of both the 0 and the $\pi$ phases in the \hbox{Nb-Fe$_{0.1}$Si$_{0.9}$-Nb}
junction has been interpreted in terms of the "vortex state" model proposed by Bulaevskii for JJs
with an insulating barrier with magnetic impurities. The adequate detailed theory which fully
describes our experimental data is currently lacking and further interactions between theory and
experiment are needed to reveal the nature of the phase-shifting effect in JJs with an insulating
paramagnetic barrier.

The authors would like to thank S. M. Frolov and M. Mihalkovi\v{c} for useful discussion. This work
is supported by the VEGA grant Agency, projects No. 2/3119/23, and 2/5130/25, by Belgian IUAP
project and by the GOA/2004/02 and the FWO Programs. O. V. acknowledges the support of the BELSPO
and ESF programme Arrays of Quantum Dots and Josephson Junctions.

\end{document}